\documentclass[10pt]{article}

\usepackage{amsmath}
\usepackage{amssymb}

\usepackage{graphicx}

\usepackage{cite}

\usepackage{color} 

\usepackage[labelfont=bf,labelsep=period,justification=raggedright]{caption}

\makeatletter
\renewcommand{\@biblabel}[1]{\quad#1.}
\makeatother

\date{}

\begin{document}

% Title must be 150 characters or less
\begin{flushleft}
{\Large
\textbf{Colocalization of coregulated genes: a steered molecular dynamics study of human chromosome 19}
}
% Insert Author names, affiliations and corresponding author email.
\\
Marco Di Stefano$^{1}$, 
Angelo Rosa$^{1,\ast}$, 
Vincenzo Belcastro$^{2}$
Diego di Bernardo$^{3,4}$
Cristian Micheletti$^{1,\ast}$
\\
\bf{1} International School for Advanced Studies (SISSA), Trieste, Italy
\\
\bf{2} Philip Morris International R\&D, Philip Morris Products S.A., Neuch\^atel, Switzerland
\\
\bf{3} Telethon Institute of Genetics and Medicine (TIGEM), Napoli, Italy
\\
\bf{4} Department of Informatics and Systems Engineering, University ``Federico II'', Napoli, Italy
\\
$\ast$ E-mail: anrosa@sissa.it; michelet@sissa.it
\end{flushleft}
% Please keep the abstract between 250 and 300 words
\section*{Abstract}
The connection between chromatin nuclear organization and gene
activity is vividly illustrated by the observation that
transcriptional coregulation of certain genes appears to be directly
influenced by their spatial proximity. This fact poses the more
general question of whether it is at all feasible that the numerous
genes that are coregulated on a given chromosome, especially those at
large genomic distances, might become proximate inside the
nucleus. This problem is studied here using steered molecular dynamics
simulations in order to enforce the colocalization of thousands of
\emph{knowledge-based} gene sequences on a model for the gene-rich
human chromosome 19. Remarkably, it is found that most ($\approx
80\%$) gene pairs can be brought simultaneously into contact. This is
made possible by the low degree of intra-chromosome entanglement and
the large number of cliques in the gene coregulatory network. A clique
is a set of genes coregulated all together as a group. The constrained
conformations for the model chromosome 19 are further shown to be
organised in spatial macrodomains that are similar to those inferred
from recent HiC measurements. The findings indicate that gene
coregulation and colocalization are largely compatible and that this
relationship can be exploited to draft the overall spatial
organization of the chromosome \emph{in vivo}. The more general
validity and implications of these findings could be investigated by
applying to other eukaryotic chromosomes the general and transferable
computational strategy introduced here.

% Please keep the Author Summary between 150 and 200 words
% Use first person. PLoS ONE authors please skip this step. 
% Author Summary not valid for PLoS ONE submissions.   
\section*{Author Summary}
Recent high-throughput experiments have shown that chromosome regions
(\emph{loci}) which accommodate specific sets of coregulated genes can
be in close spatial proximity despite their possibly large sequence
separation. The findings pose the question of whether gene
coregulation and gene colocalization are related in general. Here, we
tackle this problem using a knowledge-based coarse-grained model of
human chromosome 19. Specifically, we carry out steered molecular
dynamics simulations to promote the colocalization of hundreds of gene
pairs that are known to be significantly coregulated.  We show that
most ($\approx 80\%$) of such pairs can be simultaneously
colocalized. This result is, in turn, shown to depend on at least two
distinctive chromosomal features: the remarkably low degree of
intra-chain entanglement found in chromosomes inside the nucleus and
the large number of cliques present in the gene coregulatory
network. The results are therefore largely consistent with the
coregulation-colocalization hypothesis. Furthermore, the model
chromosome conformations obtained by applying the coregulation
constraints are found to display spatial macro-domains that have
significant similarities with those inferred from HiC measurements of
human chromosome 19. This finding suggests that suitable extensions of
the present approach might be used to propose viable ensembles of
eukaryotic chromosome conformations \emph{in vivo}.
%~198 words

\section*{Introduction}
The advent of innovative fluorescence-based techniques has provided an
unprecedented insight into the organization of eukaryotic chromosomes
during various phases of the cell cycle \cite{cremer1,pombo}. A
notable example is given by the demonstration - based on imaging
techniques - that when the tightly packed mitotic chromosomes enter
interphase they swell and occupy specific nuclear regions, aptly
termed ``territories'' \cite{cremer1}. More recently, the salient
local and global spatial properties of chromatin fibers inside these
territories have been addressed by the so-called ``chromosome
conformation capture'' techniques
\cite{dekker3c,simonis4C,dekker_betaglobin,hic,tanay2011}, which allow
for probing the cis/trans contact propensity of various chromosomal
\emph{loci}.

The recent systematic application of these experimental techniques is
providing increasing evidence that chromosomes are organized in
functionally-heterogeneous macro-domains with different molecular and
genetic composition \cite{hic,sextonHiCFly,Dixon_et_al}.

Several efforts are being spent to clarify the functionally-oriented
implications of such chromosomal organization. Towards this goal, some
of us have recently carried out a comprehensive bioinformatic survey
of data gathered in more than 20,000 gene expression profiles measured
for several cell lines in different human tissues
\cite{dibernardo}. It was thus established that genes can be grouped
into large clusters based on significant pairwise correlations (mutual
information) of their expression patterns. In addition, the matrix of
pairwise gene expression correlations displayed features qualitatively
similar to the matrix of pairwise gene contacts inferred from the HiC
\cite{hic}.

Furthermore, for various model organisms, specific sets of genes that
are systematically coexpressed were shown to be in spatial contact too
\cite{spilianakis2005,cavalli,fullwood_Nature2009}. A chief example is
provided by the human IFN-$\beta$ gene, an $\approx 800$
basepairs-long region on human chromosome 9. This gene, during virus
infection, induces colocalization and coexpression of $3$ distant
NF-$\kappa$B bound genomic \emph{loci} \cite{apostolou_thanos}.

While not all sets of coexpressed or coregulated genes are expected to
be nearby in space \cite{BystrickyPlosGen2010}, several arguments and
model calculations have consistently indicated that the simultaneous
colocalization of multiple genes can occur with appreciable
probability even when the genes are far apart along a chromosome and
in the presence of a crowded nuclear environment
\cite{CookMarenduzzoMicheletti,KepesPlosCB2010}. Indeed, it has been
argued that the cooperative colocalization of various genes can
provide a very efficient means for achieving their functional
coregulation \cite{CookScience1999,cookJMB2010}.

These considerations motivated the present numerical study where a
knowledge-based coarse-grained model of eukaryotic chromosome
19 is used to ascertain whether the large number of
coregulated gene pairs on a given chromosome can be actually
colocalized in space. The analysis therefore complements recent
efforts through which the organization of model chromatin fibers was
investigated by bringing distant regions into contact by using
attractive interactions, which either mimicked the effect of
transcription factories~\cite{KepesPlosCB2010} or 5C-based distance
restraints~\cite{BauMartiRenom}.

Our investigation, is carried out for human chromosome 19 (Chr19).
This chromosome, which is typically located at the nucleus center
\cite{hic}, was chosen because it has the highest gene density and
extensive gene expression data are available for it. By analysing the
mutual information content of thousands of such expression profiles we
identify hundreds of coregulated gene pairs for Chr19. These
coregulated gene pairs are next mapped onto a previously-validated
model for interphase chromosomes (where the chromatin filament is
coarse-grained at a resolution of $\approx 30$nm) and their pairwise
colocalization is enforced using a steered molecular dynamics scheme.
The protocol is applied to various initial chromosome configurations
where the degree of entanglement is comparable to that expected for
chromosomes {\em in vivo} (based on the crumpled-globule
interpretation of HiC data \cite{plospaper,hic}) or much higher (as in
equilibrated polymer chains). Further terms of comparisons were
obtained by randomizing the positions or pairings of the {\em loci} to
be colocalised.

Notably, for initial chromosome conformations with low entanglement,
it is found that most ($\approx 80\%$) of the coregulated gene pairs
can indeed be brought into contact and this promotes the formation of
spatial macrodomains similar to those inferred from HiC measurements
of human chromosome 19. The percentage of satisfied
colocalization constraints, and the macrodomain similarity is
dramatically reduced when the initial chromosome arrangements are
significantly entangled and when the coregulatory network is changed
by suppressing the numerous native coregulatory cliques, 
that is groups of genes all mutually coregulated.

The observed compliance of the model chromosomes towards the gene
colocalization demonstrates that bringing into simultaneous spatial
proximity most of the thousands of coregulated gene pairs for Chr19 is
physically viable. The findings are therefore consistent with the
hypothesis that coregulated genes are likely to be in contact
too. This conclusion is further supported by the fact that the spatial
macro-domains found in the constrained, steered conformations of Chr19
are well-consistent with those inferred from Hi-C data.

% Results and Discussion can be combined.
\section*{Results / Discussion}

\subsection*{Colocalization of coregulated genes in human chromosome 19}
A number of experimental studies have given the consensual indication
that various sets of coregulated genes tend to be nearby in space,
even if they are at a large genomic distance (reviewed in
Ref.~\cite{cavalli}).  Because gene colocalization is not necessary in
principle to achieve gene coregulation or coexpression (the latter
can, for instance, be induced by controlled hormone addition
\cite{BystrickyPlosGen2010}) it is not clear whether there exists a
general connection between gene coregulation and gene colocalization
and what would be the general biological implications.

In particular, two such important ramifications regard the interplay
of chromosome conformational arrangement and gene expression or
regulation.  The first issue relates to the entanglement of the long
and densely packed chromatin filaments: is their arrangement too
intricate to allow for the simultaneous colocalization of all (or
most) pairs of coregulated genes?  Secondly, in case there exists a
strong association between gene coexpression and colocalization, is it
at all feasible to use gene coexpression data as distance restraints
to pin down viable chromosome conformations?

To make progress on these standing issues we developed and used a
knowledge based numerical approach to investigate the gene
coregulation--colocalization relationship in human Chr19 using a
coarse-grained chromosome model.

Chr19 which is $\approx 60Mbp$ long, was chosen because it has the
highest gene density compared to other chromosomes \cite{chr19}. This
property reflects, in turn, in the possibility to use publicly
available gene expression data to derive knowledge based
colocalization constraints that cover extensively
Chr19.

To this purpose we started by considering $20,255$ expression
measurements for $1,278$ probesets for Chr19. As customary we shall
hereafter refer to the probesets simply as genes.  By analysing this
large pool of data using the approach described in the Materials and
Methods section, we singled out 1,487 pairs of genes which, according
to the high mutual information content of their expression profiles,
are deemed to be significantly coregulated \cite{kohane}.

Notably, the selected pairs of genes are typically far apart along the
chromosome contour. The median genomic separation of
the midpoints of the coregulated genes is as large as $25Mbp$.  

To clarify whether, and to what extent, the coregulated gene pairs can
be simultaneously colocalized we used a coarse-grained model for
chromatin filaments that has been previously shown to be capable of
accounting for the fractal-like organization observed for eukaryotic
chromosomes
\cite{grosberg,albiez2006,hic,plospaper,Vettorel2009,rev_mirny,mirny2011}.
Specifically, we adopted the model of Ref.~\cite{plospaper} where
chromatin is described as a homogeneous chain of beads with effective
diameter equal to $30 nm$ and persistence length equal to $150nm$.
Accordingly, Chr19 is described as a chain of $19,710$ beads, for a
total contour length of $\approx 591\mu m$.

\begin{figure}[!ht]
\begin{center}
\includegraphics[width=4in]{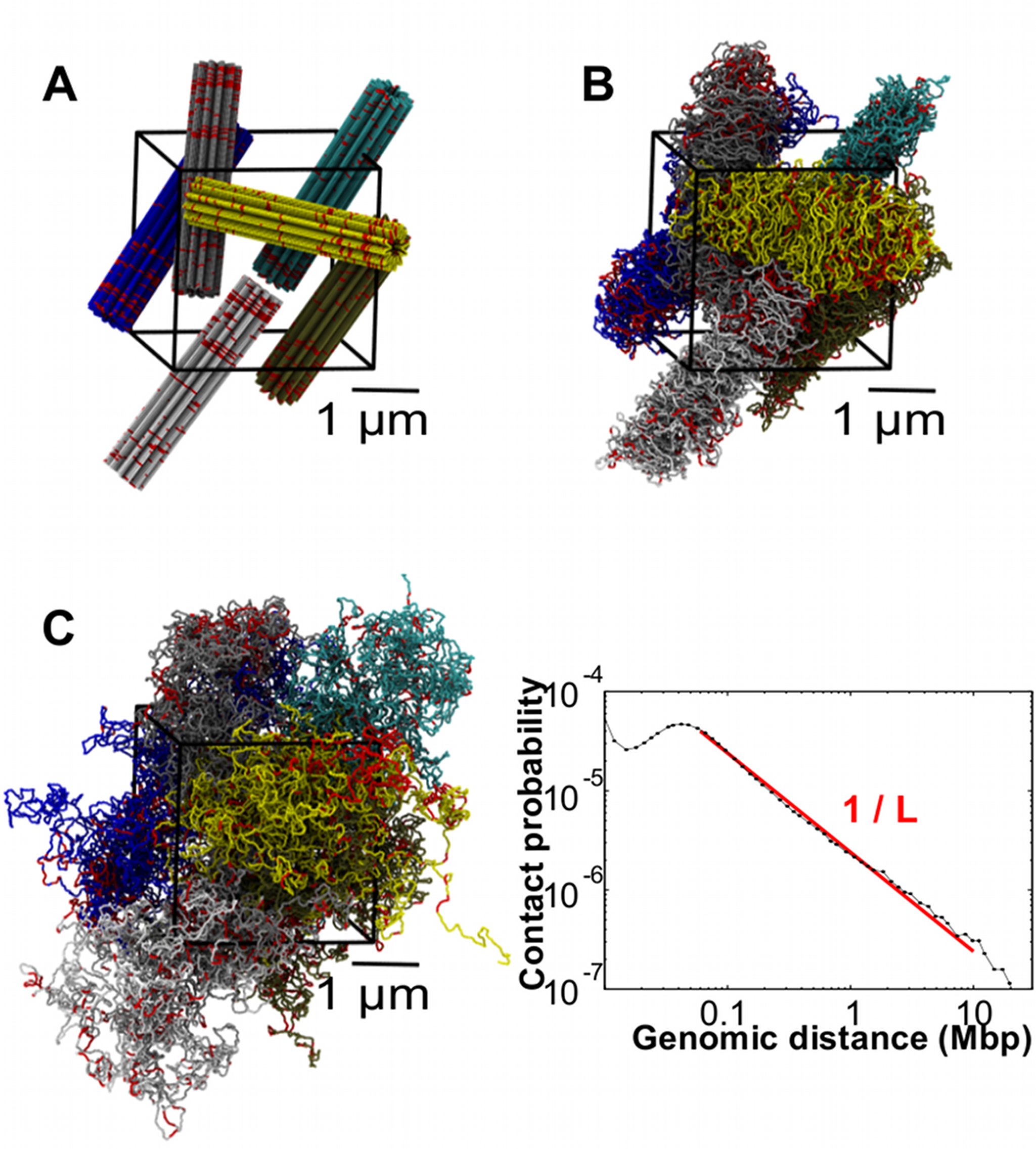}
\end{center}
\caption{{\bf Mitotic and interphase configurations of the model system chromosomes.}
(A)
Initial mitotic-like arrangement, constituted by 6 copies of model human chromosome 19.
Following ref. \cite{plospaper}, the chromatin fiber is helicoidally arranged into loops of $\approx 50$ kilo-basepairs each,
and departing radially from a central axis.
The six solenoidal arrangements were next placed in a random,
but non-overlapping manner inside a cubic simulation box of side equal to $3.0 \mu m$ and with periodic boundary conditions.
(B)
Chromosome spatial arrangement after short relaxation with a standard push-off protocol of $10^5$ MD time steps (see Methods).
(C)
Interphase-like configuration obtained by evolving the initial mitotic configuration for $10^8 t_{int}$ MD time steps
(approximately corresponding to $7$ hours in ``real-time''~\cite{plospaper}).
(\emph{Inset}) The corresponding contact probabilities between \emph{loci} of model interphase chromosomes
decay as a power law of the genomic distance, $\approx L^{-1}$,
consistent with recent experimental observations \cite{hic,bjpaper}.
In all panels, chromosome regions involved in the coregulatory network are highlighted in red.}
%should be left justified, as specified by the options to the caption 
%package.
%}
\label{fig:IniConfigs}
\end{figure}

To mimic inter-chromosome interactions in the dense nuclear
environment, we considered a system where six copies of Chr19 are
placed in a cubic simulation box (with periodic boundary conditions)
of side equal to $3 \mu m$. The overall system density is therefore
0.012 bp /nm$^3$, which corresponds to a 10\% volume fraction. Such
density matches the typical genomic one in human cell nuclei ($\approx
6 \cdot 10^9$bp in a nucleus that is $\sim10 \mu m$ in
diameter~\cite{plospaper}).  To mimic the mitotic state, each model
chromosome was initially prepared in an elongated solenoidal-like
configuration \cite{plospaper}, and the six copies were placed in a
random, but non-overlapping arrangement inside the cubic simulation
box as shown in Fig.~\ref{fig:IniConfigs}A.  To remove any excessive
intra-chain strain of the orderly designed mitotic arrangement, the
model chromosomes of Fig.~\ref{fig:IniConfigs}A were briefly evolved
with an unbiased MD protocol. The resulting relaxed mitotic
configuration is shown in Fig.~\ref{fig:IniConfigs}B.

This mitotic arrangement was further evolved for a much longer
simulation time, roughly corresponding to $7$ hours in ``real-time''
\cite{plospaper}, to obtain the fully decondensed arrangement shown in
Fig.~\ref{fig:IniConfigs}C.  Such configuration exhibits the same
power-law decay of contact probabilities versus genomic separation as
observed in HiC experiments \cite{hic,bjpaper}, see inset of
Fig.~\ref{fig:IniConfigs}C.  The model system therefore aptly
reproduces the salient experimentally-observed features of
interphase chromosomes.

After setting up the mitotic and interphase systems, we next applied a
steered molecular dynamics protocol to each of them (see Methods) to
promote the spatial proximity of regions corresponding to coregulated
gene pairs.

\begin{figure}[!ht]
\begin{center}
\includegraphics[width=4in]{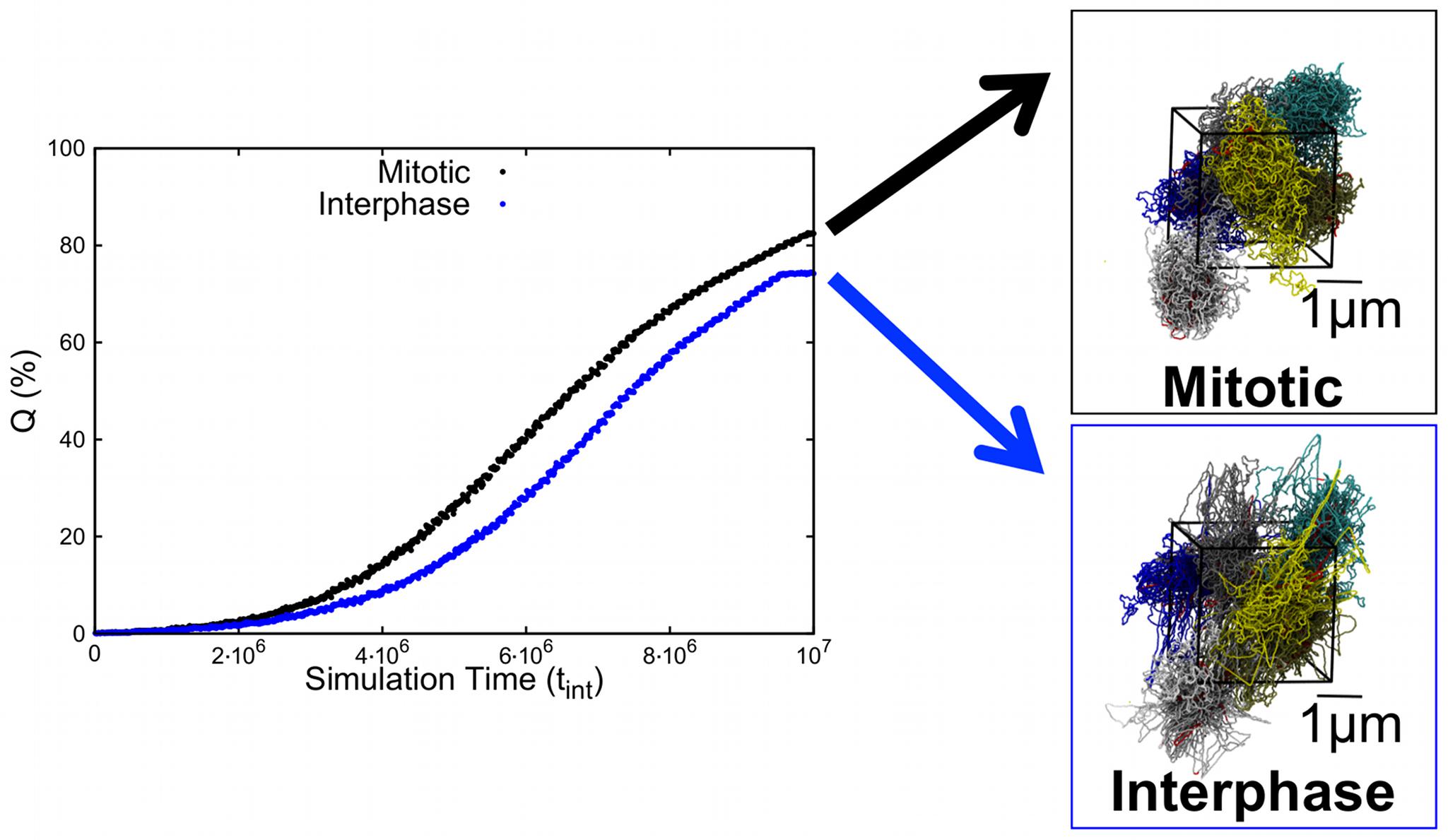}
\end{center}
\caption{ {\bf Increase of the percentage, $Q$, of Chr19 coregulated
    pairs which colocalize during the MD steering protocol.}  The two
  curves reflect different initial conditions corresponding to the
  mitotic and the interphase conformations of panels (B) and (C) of
  Fig. \ref{fig:IniConfigs}. The final configurations, corresponding
  to $Q \approx 80\%$ are shown on the right.  Chromosome regions
  involved in the coregulatory network are highlighted in red. These
  and other graphical representations of model chromosomes were
  rendered with the VMD graphical package~\cite{VMD}.}
%should be left justified, as specified by the options to the caption 
%package.
\label{fig:NativeOverlaps}
\end{figure}

The compliance of the two systems to the steering protocol is
illustrated in Fig. \ref{fig:NativeOverlaps} which shows the increase of the
percentage of target gene pairs that are successfully colocalized.

It is striking to observe that for both system it is possible to
simultaneously colocalize a very high fraction of the target pairs,
namely 80\% of them (averaged over the six chromosome copies). The
conformations reached at the end of the steering protocol are shown in
the right panels of Fig.~\ref{fig:NativeOverlaps}.

Considering the relatively-high density of the simulated system of
chromosomes and that most of the coregulated pairs lie at large genomic
distances, the results point to an unexpectedly high degree of
plasticity of the mitotic and interphase conformations, which is
presumably ascribable to their fractal-like metric properties which
keeps at a minimum the entanglement of the chromatin fiber
\cite{grosberg,albiez2006,hic,plospaper,Vettorel2009,rev_mirny,mirny2011,mirny2012}.

A second noteworthy feature of the results of
Fig.~\ref{fig:NativeOverlaps} emerges considering the diversity of the
sources used to derive the knowledge-based coregulation data.  In
fact, granted the validity of the coregulation--colocalization
hypothesis, one might have envisaged \emph{a priori} that the
chromosomal configurations corresponding to different tissues or
experimental conditions would be so heterogeneous that it would be
impossible to satisfy the cumulated set of colocalization constraints.
By contrast, the results of Fig.~\ref{fig:NativeOverlaps} demonstrate
\emph{a posteriori}, that the set of pairwise colocalization
constraints are largely mutually compatible because most of them can
be simultaneously satisfied.

The findings are therefore not only consistent with the
coregulation--colocalization hypothesis but, based on such hypothesis,
also suggest that the conformations adopted by a chromosome in various
conditions can share a common underlying pattern of colocalized genes.

\subsection*{Spatial macrodomains: comparison with data based on HiC maps}
To further characterize the overall organization of the steered
conformations shown in Fig.~\ref{fig:NativeOverlaps} we identified
their spatial macrodomains and compared them with those inferred from
the analysis of HiC data collected by Dixon {\em et
  al.}~\cite{Dixon_et_al}.

\begin{figure}[!ht]
\begin{center}
\includegraphics[width=4in]{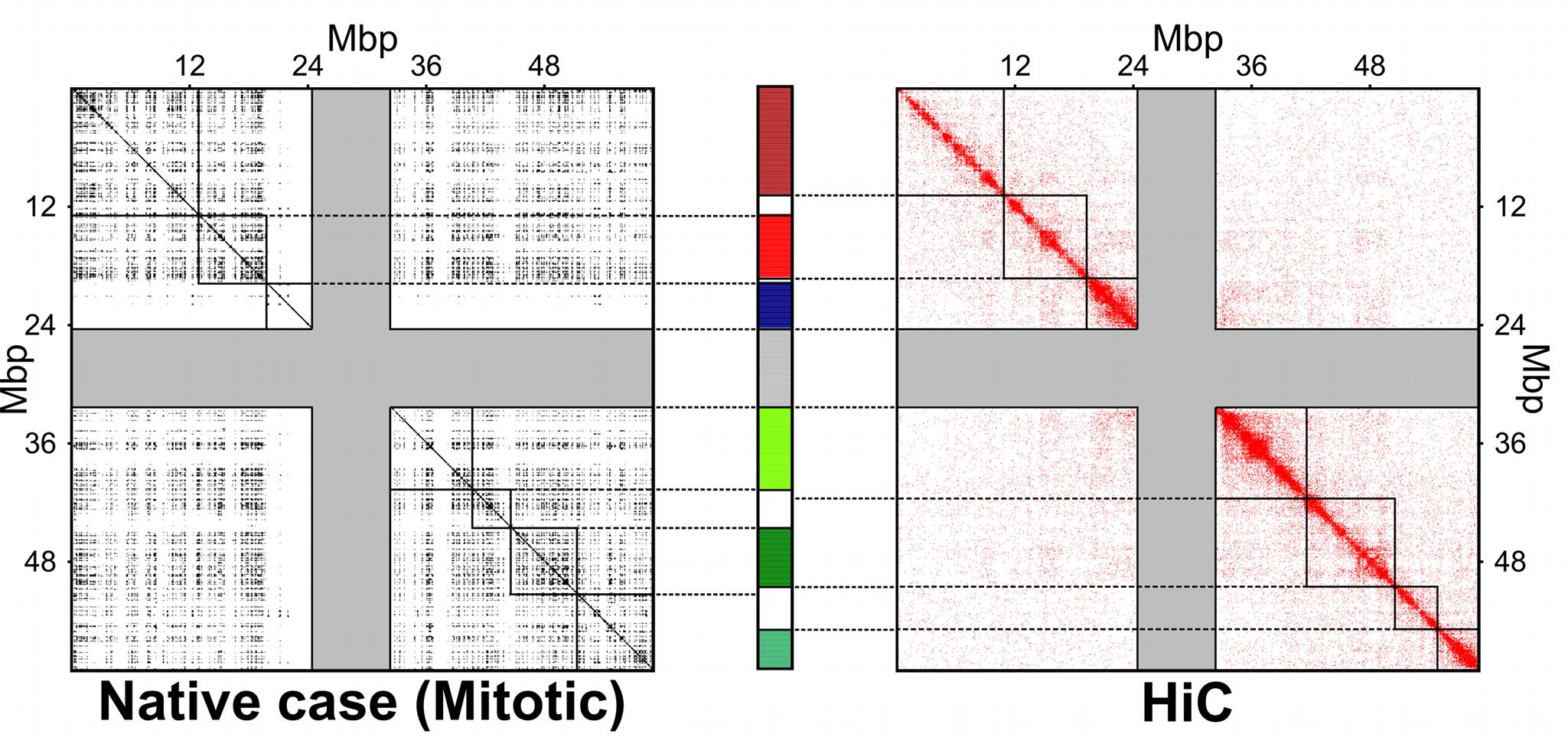}
\end{center}
\caption{ {\bf Spatial macrodomains.} The contact maps for Chr19
  obtained at the end of the steered-MD simulations and inferred from
  HiC data are shown on the left and right, respectively. The grey
  bands mark entries involving the centromere region. The boundaries of
  the $8$ principal spatial domains, identified wih a clustering
  analysis of the contact maps, are overlaid on the matrices. The
  consistency of the two macrodomain subdivisions is visually conveyed
  in the chromosome sketch at the center. The overlapping portions of
  the domain subdivisions are colored (different colors are used for
  different domains). Non-overlapping regions are shown in white,
  while the centromere region is shown in grey. The overlaping regions
  accounts for $79\%$ of the chromosome (centromere excluded).}
\label{fig:Domains}
\end{figure}

In both cases, the starting point of the analysis was the construction
of the chromosome contact map with a $60$kbp resolution, which is
commensurate with both the experimental resolution (20kbp) and the
bead equivalent contour length (3kbp). The HiC-data-based contact map
was derived from the contact enrichment values reported by Dixon {\em
  et al.}~\cite{Dixon_et_al} while the simulation-based one was
computed from the bead pairwise distances at the end of the steering
protocol (averaged over the six chromosome copies), see Methods. Both
matrices are shown in Fig.~\ref{fig:Domains}.

A clustering analysis of the contact maps was next used to subdivide
Chr19 into up to ten spatial macrodomains, each spanning an
uninterrupted chromosome stretch, and with the proviso that one domain
should cover the centromere. For both maps the consensus domain
boundaries were well-captured by the subdivision into eight spatial
domains, see Fig.~\ref{fig:consensus_domains}. The corresponding
macrodomain partitions are overlaid on the contact maps of
Fig.~\ref{fig:Domains}.

The good consistency of the domains found using HiC-based and
steered-MD contacts maps is visually conveyed by the matching colored
regions in the schematic chromosome partitioning of
Fig.~\ref{fig:Domains}. It is interesting to notice that the two
domain subdivisions consistently indicate larger domains for the upper
arm.  Quantitatively, the overlap of the two subdivisions is $0.79$,
which has a $p$-value smaller than $0.03$. This means that random
partitions of the chromosome into eight domains (one always being the
centromere) yields overlaps $\ge 0.79$ in less than $3\%$ of the
cases, see Fig.~\ref{fig:chrom_comparison}. The quantitative
comparison therefore indicates a statistically-significant consistency
of the spatial macrodomains arising in the steered chromosome
conformations and those inferred from experimental data.

\subsection*{Chromosome entanglement, regulatory network properties and gene colocalizability}
Besides the previous considerations, the results of Fig.~\ref{fig:NativeOverlaps} prompt the
question of whether, and to what extent the feasibility to colocalize a
significant fraction of the coregulated gene pairs depends on
distinctive chromosomal features, such as the spatial arrangement of
the mitotic and decondensed states or the network of coregulated 
genes.

\begin{figure}[!ht]
\begin{center}
\includegraphics[width=4in]{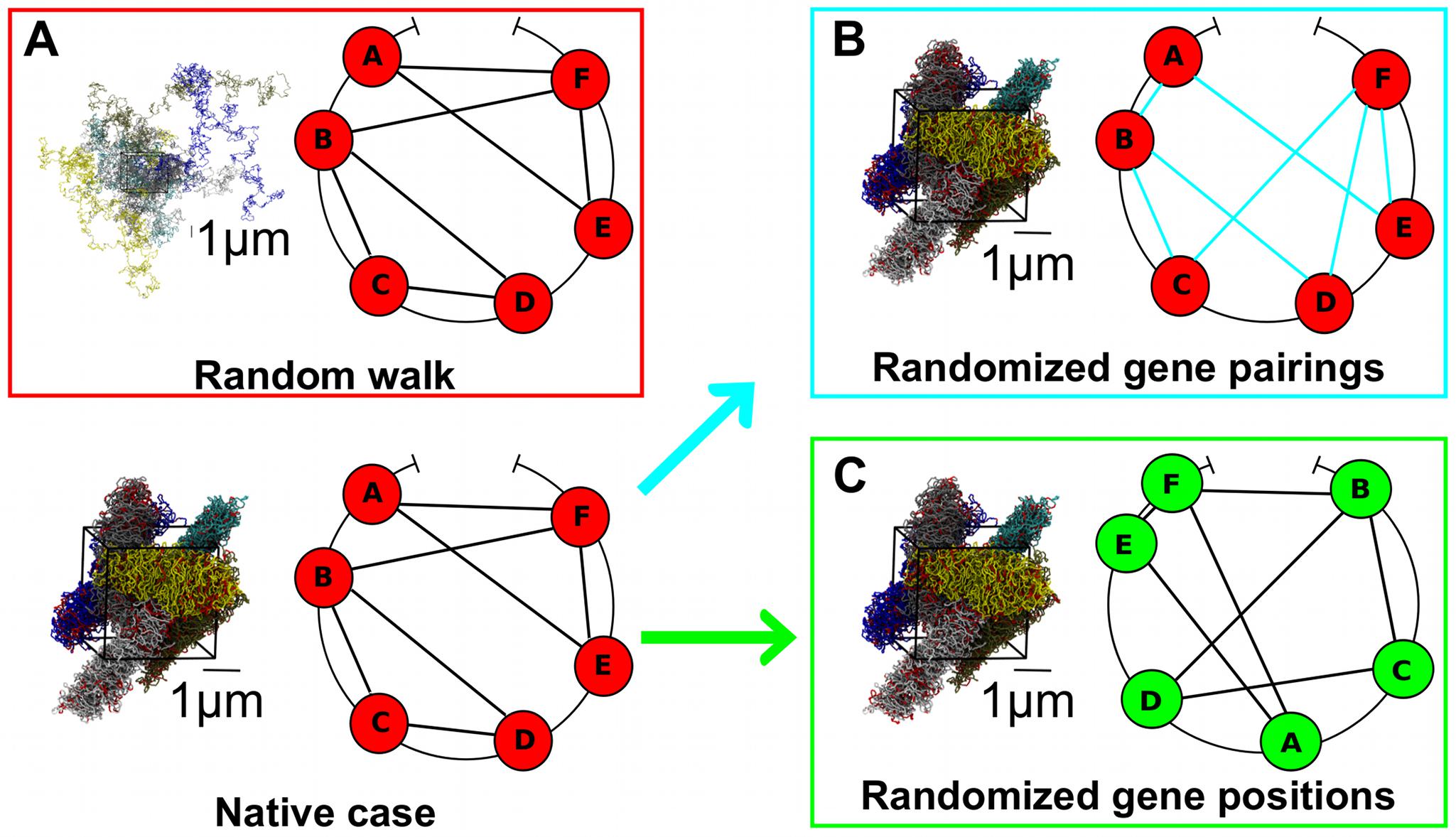}
\end{center}
\caption{
{\bf Variant systems subjected to the MD steering protocol.}
(A)
Initial configuration of 6 random-walk like chains the linear size the model chromosome 19.
(B)
Model chromosomes were initially arranged as in the mitotic-like configuration of Fig. \ref{fig:IniConfigs}B, but
the pairings between genes were randomized. The randomization preserved the number of pairs that each probeset takes part to.
(C)
Model chromosomes were initially arranged as in the mitotic-like configuration of Fig. \ref{fig:IniConfigs}B, but the
gene positions along the chromosome were randomized. The randomization preserved the native pairings of the genes.
In all panels chromosome regions involved in the native or randomized coregulatory network are highlighted in red.
For all the three systems considered the same physical conditions of fiber density, stiffness and excluded volume interactions of the original system apply.
%should be left justified, as specified by the options to the caption 
%package.
}
\label{fig:IniVariants}
\end{figure}

To address these issues we re-applied the steering protocol starting from 3 different initial conditions,
which correspond to specifically designed variants of the model chromosomes.
Specifically, the three systems are:
\begin{enumerate} 
\item
A \emph{random-walk-like} chromosome arrangement as shown and described in Fig.~\ref{fig:IniVariants}A.
\item
A mitotic-like spatial arrangement but with \emph{randomized gene pairings}, see Fig.~\ref{fig:IniVariants}B.
The chromosome spatial configuration is the same as in Fig.~\ref{fig:IniConfigs}B, but the
native 1,487 coregulatory pairings between the 412 selected genes have been randomly reshuffled.
The number of pairings that each selected gene takes part to in the reshuffled network is the same as the native coregulatory network.
\item
A mitotic-like spatial arrangement but with \emph{randomized gene positions}, see Fig.~\ref{fig:IniVariants}C.
As in case 2 above, the chromosome spatial configuration is again the same as in Fig.~\ref{fig:IniConfigs}B,
but the positions of the 412 genes involved in the native coregulatory network are randomly assigned along the chromosome (except for the centromeric region).
The repositioned genes inherit the native coregulatory pairings.
\end{enumerate}

\noindent As for the native network of target gene pairs, we report on
the properties measured at the end of the steering protocol after
averaging them over the six chromosome copies in the simulation cell.

We stress that the three variants are prepared so to preserve
the native overall density, number of coregulated genes and also the
number of coregulated pairs to which a selected gene takes part to. 
They nevertheless present major differences which allow for probing the
impact of different system properties on gene  ``colocalizability''.

\begin{figure}[!ht]
\begin{center}
\includegraphics[width=4in]{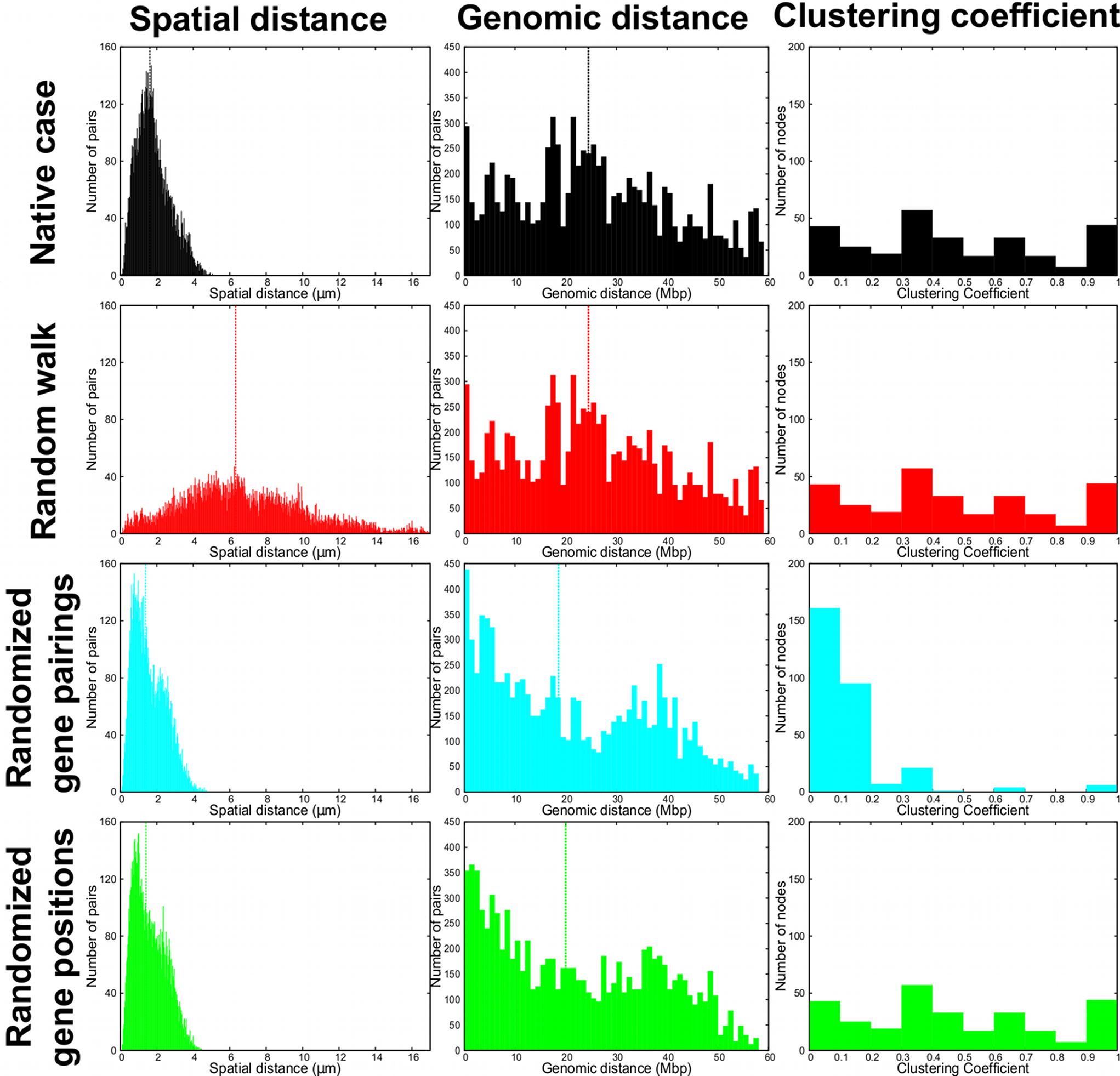}
\end{center}
\caption{
{\bf Summary of the structural properties of the native system (Fig. \ref{fig:IniConfigs}B) and its three variants (Fig. \ref{fig:IniVariants}).}
(First column)
Distribution of the \emph{spatial} distances between steered \emph{loci}. The distribution of the random-walk-like is broader than the native case one. The randomized position and randomized pairs cases have instead a similar distribution with respect to the native case.
(Second column)
Distribution of the \emph{genomic} distances between steered \emph{loci}.
(Third column)
Clustering coefficients (see Methods) of the corresponding networks of pairings between steered \emph{loci}.
Dashed lines correspond to the median values.
The results are cumulated over all $6$ chromosome copies in the simulation box.}
\label{fig:NativeVSAll}
\end{figure}

In particular, the random-walk-like arrangement has a much higher
degree of intra- and inter-chain entanglement than all other
arrangements, as illustrated by the much wider distribution of gene
pairwise distances in the initial configuration, see
Fig.~\ref{fig:NativeVSAll}.  For randomly-paired and
randomly-repositioned genes, instead, the distributions of genomic
distances of the target genes to be paired is similar to the native
one. This is clearly shown by the distributions in
Fig.~\ref{fig:NativeVSAll}. However, the same figure clarifies that
the two randomized cases differ markedly from the native one for the
clustering coefficient. The clustering coefficient captures the degree
of cooperativity of the (putative) coregulatory network in that it
measures how frequently two genes that are both coregulated with a
third one, are themselves coregulated too. The inspection of the
rightmost graphs in Fig.~\ref{fig:NativeVSAll} therefore indicates
that the clustering coefficient distribution of the randomly-paired
system is shifted towards much smaller values than the others, which
all inherit the native pairings network. This fact indicates that the
clustering coefficient of the native network is significantly larger
than random. This implies that genes can frequently interact
concertedly in groups of three or more.

\begin{figure}[!ht]
\begin{center}
\includegraphics[width=4in]{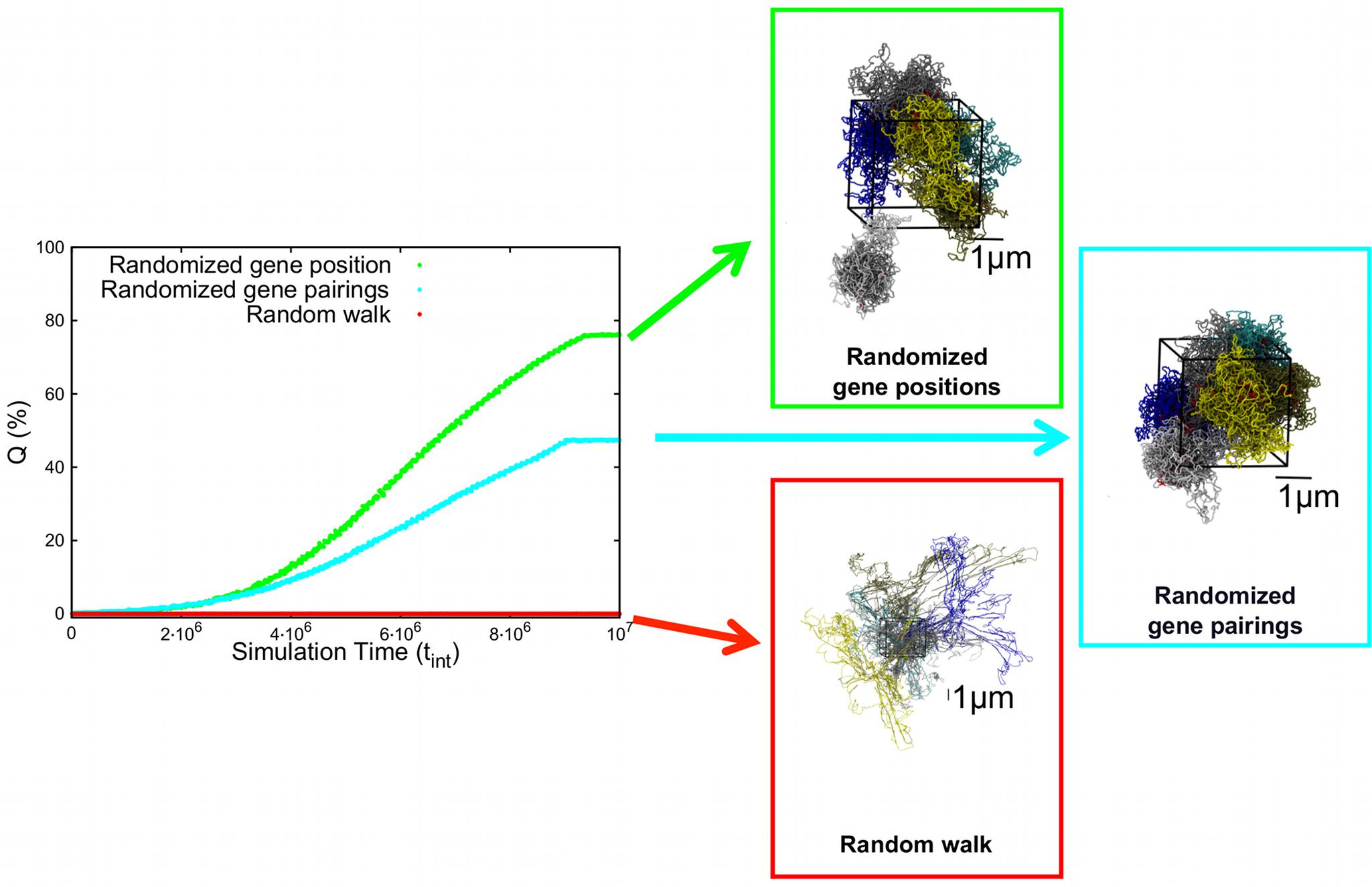}
\end{center}
\caption{
{\bf Increase of the percentage, $Q$, of Chr19 coregulated pairs which colocalize during the MD steering protocol,
for the three variants of the native systems.}
The configurations reached at the end of the steering protocol are shown on the right.
Chromosome regions that take part to the pairs of loci to be colocalized are highlighted in red.
%should be left justified, as specified by the options to the caption 
%package.
}
\label{fig:IniVariantsResults}
\end{figure}

The results of the steering protocol applied to the three system
variants are shown in Fig.~\ref{fig:IniVariantsResults}.
The data indicate that:
(i)
for random-walk-like chromosomes only a minute fraction ($<$1\%) of the target contacts can be satisfied;
(ii)
for randomly-paired genes about $47$\% of the gene pairs can be colocalized, while
(iii)
for randomly-repositioned genes about $75$\% of the gene pairs can be colocalized,
similarly to the native case (Fig. \ref{fig:NativeOverlaps}).

These findings provide valuable clues for interpreting the high degree
of ``colocalizability'' of coregulated genes observed in
Fig.~\ref{fig:NativeOverlaps} for the mitotic and interphase
arrangements.

In particular, the very low asymptotic value of the percentage of
successfully colocalized gene pairs for the random-walk-like system
clarifies that the low intra- and inter-chromosome entanglement of
both the mitotic and decondensed configurations is crucial for
bringing into contact the coregulated gene pairs.

Furthermore, the comparison of the randomly-paired and
randomly-repositioned gene cases shows that the connectivity
properties of the native coregulatory network appear even more
important than the detailed positioning of the coregulated genes along
the chromosomes. In fact, the randomly-repositioned genes -- which
retain the same clustering coefficient of the native coregulatory
graph -- have the same high degree of colocalizability of the native
system. By converse, the low clustering coefficient of the
randomly-paired gene case -- corresponding to a significant disruption
of the original network -- reflects in an appreciably lower value of
percentage of successfully colocalised gene pairs. It is also worth
noticing that, in all cases, a significant fraction of gene pairs
brough in contact are at large genomic distances ($>20Mbp$), see
Fig.~\ref{fig:unsatisfied_pairings}.

Finally, the network randomization effects on the spatial organization
of the steered conformations was addressed by measuring the overlap of
their spatial macrodomains with those established from HiC data. We
recall that for chromosome subdivisions into eight macrodomains, the
native case overlap was $0.79$. For the randomized gene positions and
randomized gene pairings we instead observe the lower values $0.73$
and $0.63$, respectively. These values clearly have a much lower
statistical significance than the native case; their $p$-values being
respectively $0.113$ and $0.490$, see
Fig.~\ref{fig:chrom_comparison}. Their non-significant similarity with
the reference, HiC-data based macrodomain subdivisions underscores the
randomized, non-native constraints result in appreciably-different,
and less realistic, chromosomal features.

\subsection*{Summary and conclusions}
Recent experimental advancements have provided unprecedented insight
into the occurrence of concerted transcription of multiple genes.
In particular, it was reported that the chromatin fiber can rearrange so that
genes, concertedly transcribed upon activation, are found nearby in space too.

Because of its important ramifications, the possible existence of a
general relationship between gene coregulation and gene
colocalization, the so called ``gene-kissing'' mechanism
\cite{spilianakis2005,cavalli}, is a subject of very active research.
 
This standing question was addressed here numerically by carrying out
molecular dynamics simulations of a knowledge-based coarse-grained
model of human chromosome 19. The model consisted of a coarse-grained
representation ($30nm$ resolution) of the chromatin fiber complemented
by the knowledge-based information of the loci corresponding to
($\approx 1500$) coregulated gene pairs. These pairs were identified
from the analysis of extensive sets of publicly-available gene
expression profiles. To mimic the crowded nuclear environment, we
considered a system where several copies of the model chromosome 19
were packed at typical nuclear densities. The colocalization of the
coregulated gene pairs was finally imposed by applying a steered
molecular dynamics protocol.

It was found that most ($\approx 80\%$) of the coregulated pairs could
be colocalized in space when the steering protocol was applied to
chromosomes initially prepared in mitotic-like and interphase-like
arrangements, see Fig.~\ref{fig:NativeOverlaps}. Notably, the pattern
of intra-chromosome contacts established for the steered conformations
exhibited significant similarities with that of experimental contact
propensities \cite{hic,tanay2011} of chromosome 19. Furthermore, the
overall chromosomal organization into spatial macrodomains showed
significant similarities with that inferred from experimental HiC
data.

By converse, the percentage of colocalized target pairs decreased
substantially (or vanished altogether) when the system was initially
prepared in a random-walk like arrangement, or if the genes to be
colocalized were randomly paired or displaced along the
chromosome. Likewise, the macrodomain organization of these
alternative systems was found to be much less similar to the
HiC-data-based one.

The present findings allow to draw several conclusions. First, the
data in Fig.~\ref{fig:NativeOverlaps} demonstrate that, even in a
densely packed system of mitotic or interphase chromosomes it is
physically feasible to achieve the simultaneous colocalization of a
large number of pairs of loci that can be very far apart along a
chromosome. This result is therefore well compatible with the
gene coregulation--colocalization hypothesis. In fact, the findings
can be read as adding support to the hypothesis in consideration of
the fact that if no meaningful relationship existed between
coregulation and colocalization one might have expected the
unfeasibility of bringing into simultaneous contact so many
coregulated pairs.

The much poorer compliance of alternative systems (random-walk-like
chromosome conformations, randomized gene pairings and positions) to
the steering protocol provides valuable insight into the native
chromosomal properties that allow for gene colocalization.

The first and most important property is the low degree of
entanglement that mitotic or interphase chromosomes are known to have
compared to equilibrated polymer solutions of equivalent density
\cite{grosberg,albiez2006,hic,plospaper,Vettorel2009,Marenduzzo:2010:J-Phys-Condens-Matter:21399272,rev_mirny,mirny2011,mirny2012}. The
second property is that the number of gene cliques that is present in
the native gene regulatory network of chromosome 19 is much higher
than for the equivalent random network. In this respect it is worth
pointing out that the atypically large number of cliques found in
biological regulatory networks has also been observed and pointed out
in different contexts and for a different set of chromosomes
\cite{gingeras}.

To further validate this conclusion we considered an additional target
network for the steered-MD simulations. This network was obtained by a
partial randomization of the native gene pairings and its average
clustering coefficient was $30\%$, which is intermediate to the native
one ($47\%$) and the fully-randomized case ($12\%$) discussed
previously. As shown in Fig.~\ref{fig:CC_overlap}, $64\%$ of the
target colocalization constraints were satisfied. This value is
intermediate between the native and fully-randomized case ($82\%$ and
$47\%$, respectively) and hence supports the existence of a meaningful
correlation between gene colocalizability and the regulatory network
cliquishness.

In perspective, because the computational strategy employed here is
formulated in a general and transferable way, it would be most
interesting to apply it to other eukaryotic chromosomes for which
extensive co-regulatory data is available. This could clarify the more
general validity of the gene coregulation-colocalization relationship
as well as the broader implications of using it (possibly with other
knowledge-based constraints\cite{zimmer2005,BauMartiRenom,alber}), for
charting the spatial organization of eukaryotic chromosomes, and
possibly of systems of chromosomes.

% You may title this section "Methods" or "Models". 
% "Models" is not a valid title for PLoS ONE authors. However, PLoS ONE
% authors may use "Analysis"

\section*{Materials and Methods}

\subsection*{Coregulated gene pairs on Chr19}
To identify the set of significantly coregulated gene pairs on Chr19
we processed a set of $20,255$ expression profiles of human probesets
measured in $591$ distinct microarray experiments. The gene expression
profiles, which were all measured on HG-U133A Affymetrix chip, pertain
to different human cell types and tissues in various experimental
conditions. This extensive dataset was recently compiled and curated
by some of us~\cite{dibernardo} starting from the public ArrayExpress
database \cite{arrayexpress}.

The analysis was restricted to the set of 1,278 probesets which
exclusively target a single sequence (i.e. an uninterrupted stretch)
of chromosome 19. Next, to perform a robust comparison between the
differently normalized gene expression profiles we coarse-grained all
expression levels to one of three discrete states only: low, medium
and high, as done in Ref.~\cite{dibernardo}. For each possible
probeset pair, $I$ and $J$, we next computed the mutual
information~\cite{dibernardo} content (MI) of the expression profiles:
\begin{equation}
\label{eq:mi}
{\rm MI}_{IJ} = \sum_i \, \sum_j \, \pi_{ij} \, \ln \left ( \frac{\pi_{ij}}{\pi_{i+} \, \pi_{+j}}  \right )
\end{equation}
where $i$ [$j$] runs over the three coarse-grained expression levels
for probeset $I$ [$J$]. In Eq. \ref{eq:mi}, $\pi_{ij}$ is the joint
probability that, in a given experiment, the expression levels $i$ and
$j$ are respectively observed for probesets $I$ and $J$, while the
quantities $\pi_{i+} \, = \, \sum_j \pi_{ij}$ and $\pi_{+j} \, = \,
\sum_i \pi_{ij}$ are the probabilities to observe expression level $i$
[$j$] for probeset $I$ [$J$] (marginal probabilities). The MI thus
provides a statistically-founded measure of how the gene expression
\emph{pattern} for gene $I$ is predictable assuming the knowledge of
another \emph{pattern} $J$ (or, vice versa).

\begin{figure}[!ht]
\begin{center}
\includegraphics[width=4in]{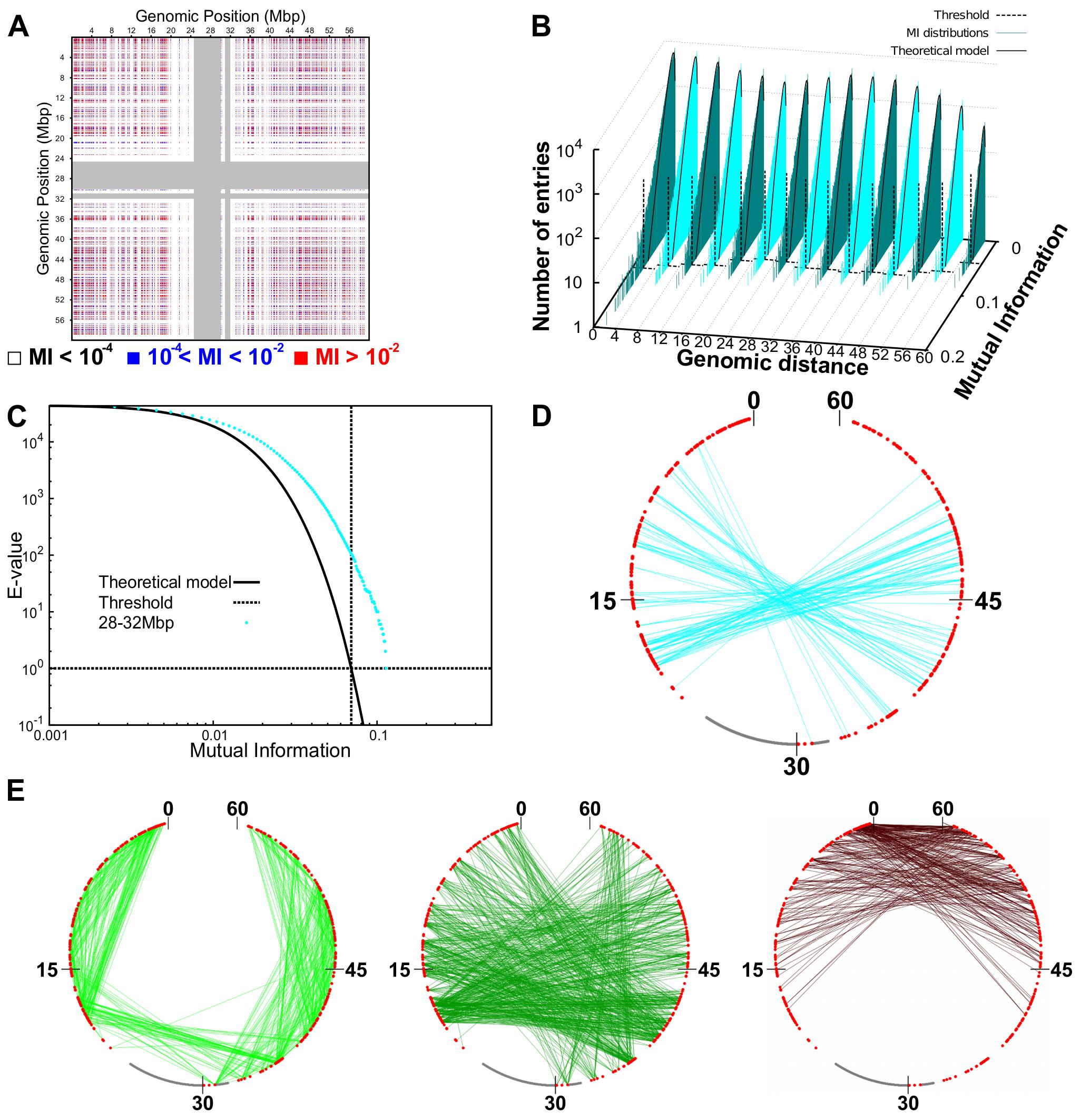}
\end{center}
\caption{
{\bf Statistical analysis of mutual information.}
(A)
Mutual information values for any pairs of probesets on Chr19.
The middle point of each probeset identifies its position along the chromosome.
The gray stripes correspond to the centromere.
(B)
Histograms of values of mutual information for pairs of probesets located at various intervals of their genomic separation.
The black lines correspond to fitting the histograms with the theoretical (null case) MI distribution \cite{Goebel}.
The vertical black dashed lines correspond to the estimated threshold values (see next and main text).
(C)
Example of E-value (expected number of false positives) distribution
for probeset pairs located at genomic separation in the range $28-32$Mbp. 
The threshold is the value of mutual information at which the E-value is equal to $1.0$.
For different genomic separations, analogous curves were obtained.
(D)
Network of coregulated pairs of genes at $28-32$Mbp separation.
The analysis illustrated in (C) singles out significantly-high values of Mutual Information. 
These contributions corresponds to connections (\emph{cyan links}) between coregulated gene pairs (\emph{red dots}).
The scale is in $\mu m$.
(E)
Networks of coregulated pairs of loci used to fix
the spatial constraints between corresponding regions of the model chromosomes.
For the seak of clarity, the whole network has been represented as three sub-networks
for pairs of loci at genomic separations of 0-20Mbp (\emph{left}), 20-40Mbp (\emph{middle}) and 40-60Mbp (\emph{right}), respectively.
%should be left justified, as specified by the options to the caption 
%package.
}
\label{fig:MutInfo}
\end{figure}

To single out the pairs of probesets with statistically-significant
coexpression we proceeded according to the procedure described below
and summarized graphically in Fig.~\ref{fig:MutInfo}.

First, to account for the expected dependence of gene coregulation on
genomic distance, we subdivided the probeset pairs in $15$ groups. The
first, second, etc. group gathered pairs of probesets whose central
bases had a genomic distance falling in the intervals 0-4Mb, 4-8Mb,
etc. Next, for each group we fitted the histogram of the pairwise MI
values, with the analytical expression $f \left( x \right)=a\, x\,
e^{-b\,x}$ which is known to approximate well the distribution of MI
values expected for two random variables (expression of the two genes)
assuming $3$ possible distinct values (low, medium and high)
\cite{Goebel}. In the previous expression $x$ is the mutual
information and $a$ and $b$ are the free fitting parameters.

The comparison with the reference, null distribution is used to define
the Mutual Information threshold above which at most one false-positive
entry is expected to occur.  All probeset pairs exceeding this stringent
MI threshold were retained (see Fig.~\ref{fig:MutInfo}C). 

The number of selected pairs for each bin ranged from $59$ to $334$,
for a total of $1,991$ probe pairs.  It should be noted that several
of these pairs involve chromosome regions that are highly overlapping
and are hence degenerate (or nearly degenerate).  To eliminate this
redundancy, we grouped together the pairs of coregulated probesets
that assure the coregulation of regions, whose central beads are
separated by less than $300 nm$ (which corresponds to the chromatin
fiber statistical (Kuhn) length \cite{plospaper}). For each of these
groups, we retained only the pair with the largest MI value. This
filtering procedure returned 1,487 non-degenerate probeset pairs, that
involved $412$ probesets (native case).  As customary, the significant
degree of coexpression of such pairs was deemed indicative of their
coregulation \cite{kohane}.

\emph{Randomized cases}. Besides the ``native case'', in which the
gene pairs to colocalize are obtained from coregulatory network of
Chr19, we considered another non-native set of target gene pairs. As
described hereafter, these alternative sets were generated by
randomizing the native gene pairing network while preserving various
overall network properties.
\begin{enumerate}
\item \emph{Randomized pairings}. The 1,487 native pairings between
  the considered set of $412$ probesets were randomly reshuffled while
  preserving the native number of pairings for each gene.  This
  alternative set of probeset pairs is obtained by applying the
  iterative randomization method described in
  ref.~\cite{JMBmicheletti}. The asymptotic fraction of randomized
  gene pairs matching the native ones is $\approx 10\%$.
\item \emph{Randomized positions}. The set of $412$ native probesets
  are randomly repositioned along the contour length of the
  chromosome, but the target gene pairings are kept the same as the
  native ones. Gene repositioning in the centromeric region (which is
  mostly void of genes) was disallowed.
\end{enumerate}

The feasibility to colocalize in space the $1,487$ pairs of probesets
was explored using the coarse-grained model chromosome and the
steering molecular dynamics protocol described in the following
subsections.

\subsection*{The chromosome polymer model}
A system of densely packed chromosomes was modelled at a resolution of
$30 nm$. Specifically, we considered $n = 6$ model chromosomes packed
at the typical nuclear density of $\approx 0.012bp/nm^3$.  Each of the
six chromatin fibers was described as a chain of $N=19,710$ beads with
diameter $\sigma = 30nm$, which corresponds to the total contour
length $L_c = 59.13Mbp$ of human chromosome 19. Each bead therefore
represents $\approx3,000$ basepairs \cite{30nm}.

The potential energy of each chain is written as,
\begin{eqnarray}
{\cal H}_{intra} & = & \sum_{i=1}^N [ U_{FENE}(i, i+1) + \notag \\
                 &   & U_{br}(i, i+1, i+2) +\\
                 &   & \sum_{j=i+1}^N U_{LJ}(i,j) ] \notag
\end{eqnarray}
where $i$ and $j$ run over the bead indices and the three terms
correspond to the FENE chain-connectivity
interaction\cite{kremer_jcp}, the bending energy, and the repulsive
pairwise Lennard-Jones interaction.  The three energy terms are
parametrized as in previous studies of coarse-grained
chromosomes~\cite{plospaper,bjpaper}. Specifically,
\begin{equation}\label{eq:fenepot}
U_{FENE}(i,i+1) = \left\{
\begin{array}{l}
- \frac{k}{2} \, R^2_0 \, \ln \left[ 1 - \left( \frac{d_{i,i+1}}{R_0} \right)^2 \right], \, d_{i,i+1} \leq R_0\\
0, \, d_{i,i+1} > R_0
\end{array}
\right.
\end{equation}
\noindent where $d_{ij}$ is the distance of the centers of beads $i$
and $j$, $R_0=1.5 \sigma$, $k=30.0 \epsilon / \sigma^2$ and the
thermal energy $\kappa_B\, T$ equals $1.0 \epsilon$ \cite{kremer_jcp}.
$U_{FENE}$ ensures the connectivity of the chain, i.e. the centers of
two consecutive beads must be at a distance about equal to their
diameter.  
The bending energy has instead the standard Kratky-Porod
form (discrete worm-like chain):
\begin{equation}\label{eq:stiffpot}
U_{br}(i, i+1 ,i+2) = \frac{K_B\, T \, \xi_p}{\sigma}  \left (1 - \frac{{\vec d}_{i,i+1} \cdot  {\vec d}_{i+1,i+2}}{d_{i,i+1} \, d_{i+1,i+2}} \right )
\end{equation}
where $\xi_p=5 \sigma=150nm$. $U_{br}$ ensures that the chain of beads
bends over contour lengths the size of the persistence length $\xi_p$
to model the experimental rigidity of the chromatin fiber
\cite{bystricky}.

Finally, the excluded volume interaction between distinct beads,
including consecutive ones, corresponds to a purely repulsive
Lennard-Jones potential:
\begin{equation}\label{eq:ljpot}
U_{LJ}(i,j) = \left\{
\begin{array}{l}
4 \epsilon [(\sigma/d_{i,j})^{12} - (\sigma/d_{i,j})^6 + 1/4], d_{i,j} \leq \sigma 2^{1/6}\\
0, d_{i, j} > \sigma 2^{1/6}
\end{array}
\right. 
\end{equation}
This repulsive interaction controls the inter-chain excluded volume too:
\begin{equation}
{\cal H}_{inter} = \sum_{I=1}^{n-1} \sum_{J=I+1}^n U_{LJ}(i,j)
\end{equation}
where $n$ is the number of chains in solution and the index $i$ $[j]$
runs over the beads in chain $I$ $[J]$. $U_{LJ}$ ensures that any two
regions along the same chain or on different chains cannot pass
through each other. In this way, intra- and inter-chain topology is
preserved.

\subsection*{Simulation details}
The LAMMPS molecular dynamics software package \cite{lammps} is used
to integrate the system dynamics at constant temperature and
volume. The integration time step was set equal to $t_{int} = 0.012
\tau_{MD}$, where $\tau_{MD} = \sigma (m/\epsilon)^{1/2}$ is the
Lennard-Jones time and $m$ is the bead mass which was set equal to the
LAMMPS default value.  Periodic boundary conditions apply.

The ``native case'' system was evolved from three different starting
conditions shown in Fig.~\ref{fig:IniConfigs}: \emph{mitotic,
  interphase} and \emph{random arrangements}, whereas the
\emph{randomized cases} systems were evolved from the \emph{mitotic}
one.

\emph{Steered Molecular Dynamics protocol.}  
The colocalization of the $1,487$ coregulated genes was attempted by
using a steered molecular dynamics protocol which progressively
favoured the spatial proximity of the pairs of genes in each of the
six model chromosomes.

Specifically, for each pair of selected genes, $A$ and $B$, we added
to the system energy an harmonic constraint,
$$
U_{harm} = \frac{1}{2} k \left( t \right) d_{A,B} ^ 2
$$
\noindent where $d_{A,B}$ is the distance of the centers of mass of
the chromosome stretches (mapped onto the discrete beads using the
Affymetrix annotation table~\cite{affymetrix}) covered by the two
genes. The stiffness of the harmonic constraint was controlled by the
time-dependent parameter $k(t)$. The latter is ramped linearly in time
from the initial value $k \left( t=0 \right) = 0.001 \epsilon /
\sigma^2$ up to the value $k \left( T_{end} \right) = 16.384 \epsilon
/ \sigma^2$. The total duration of the steered dynamics was $T_{end} =
10^7 t_{int}$. This protocol favours the progressive reduction of the
width of the distribution of probeset distances from the initially
generous value of $\approx 50 \sigma$ (see~\ref{fig:NativeVSAll}) down
to $\approx 0.4 \sigma$.  The simultaneous application of the $1,487$
constraints to each of the six chromosomes, which clearly are not
necessarily compatible \emph{a priori}, was implemented using the
PLUMED plugin for LAMMPS~\cite{plumed}.  The protocol is sufficiently
mild that no crossings of the chains should occur. This was checked by
running the steering protocol on a circularized variants of the
mitotic conformation shown in Fig \ref{fig:IniConfigs}A, and checking
that the initially unknotted topological state is maintained
\cite{PTPS_tubiana_2011}.

\subsection*{Order parameters}
To monitor the progress of the steered molecular dynamics simulations
and to characterize the salient properties of the resulting
configurations we computed two order parameters, namely the
\emph{percentage of coregulated pairs that are colocalized} and the
\emph{clustering coefficient} of the coregulated pair graph. The two
parameters are defined hereafter.

\begin{itemize}
\item 
  The \emph{percentage of coregulated pairs that are colocalized},
  $Q$, is calculated as:
  \begin{equation}\label{eq:overlap}
    Q = \frac{1}{G} \sum_{(A,B)} \Theta \left( r_c - d_{A,B} \right) \times 100 \ .
  \end{equation}
  In the above expression, the sum runs over the coregulated pairs of
  genes, $A$ and $B$ which are in total $G=8,922$ (i.e. 1,487 for each
  of the six chromosome copies), $d_{A,B}$ is the distance of their
  centers of mass.  $\Theta \left( x \right)$ is the Heaviside step
  which takes a value of 1 if $x>0$ and 0 otherwise. $\Theta$ is used
  to restrict the sum to those gene pairs that are at distance within
  the contact range, $r_c=120nm$. This cutoff distance was chosen
  because it is about equal to the typical size of a ``transcription
  factory'' \cite{cookJMB2010}.

\item 
  The \emph{clustering coefficient}, $CC$, is used to characterize
  connectivity properties of graphs. In the present case the graph of
  coregulation of pairs of genes. Each gene is represented by a node
  in the graph. Pairs of coregulated genes are represented by a link
  connecting the two corresponding nodes.

  The clustering coefficient of the individual $i$th node in the graph
  is defined as \cite{strogatz,kaiser}
  \begin{equation}\label{eq:lcc}
    c_i = \frac{\Gamma_{i}}{\gamma_i \, (\gamma_i-1)}
  \end{equation}
  \noindent where $\gamma_i$ is the number of neighbours of $i$ while
  $\Gamma_{i}$ is the number of distinct links between the neighbours
  of node $i$. The clustering coefficient per node, $c_i$, is clearly
  defined only for nodes with at least two neighbours. The clustering
  coefficient of the whole graph is obtained by averaging $c_i$ over
  all nodes with $\gamma_i \ge 2$. The clustering coefficient
  provides a measure of the incidence of cliques of size $3$
  (``triangular linkages'') in the graph.
\end{itemize}

\subsection*{Identification of spatial macrodomains}

The overall spatial organization of Chr19 was encoded in a binary
contact matrix, $C$, with a $60kbp$ resolution. The generic matrix
entry $C_{i,j}$ takes on the value $1$ or $0$ according to whether the
$i$th and $j$th 60kbp-long segments (equivalent to $20$ beads) are in
spatial proximity or not. The recent high-resolution HiC measurements
of Dixon et al.~\cite{Dixon_et_al} were used to derive the
experimental, reference contact map.  Specifically, for every
significant HiC entry (i.e. normalized contact enrichment $\ge 1$) the
corresponding contact-matrix elements were set equal to $1$. The
resulting HiC-based contact map is sparse in that only $5\%$ of its
entries are non-zero. For an equal footing comparison, we next
populated the theoretical contact maps by considering in spatial
contacts (entries equal to 1) only the top $5\%$ $60kbp$-strands
ranked for increasing average distance. The distance average is taken
over the six Chr19 copies at the end of the steering protocol.

A clustering analysis of the contact maps was next used to subdivide
Chr19 into up to ten spatial macrodomains. Each domain spans an
uninterrupted stretch of the chromosome and one domain always matches
the centromer region. Following the K-medoids clustering
strategy~\cite{hastie_kmedoids} the optimal domain partitioning was
identified by minimizing the total intra-domain
dissimilarity. Quantitatively, the internal dissimilarity of one
specific domain, covering the chain interval $i$ to $j$ is measured
as:
  \begin{equation}
    \Delta = \sum_{l=i}^j (1- C_{l, r})\ ,
  \end{equation}
  \noindent where $C$ is the contact map and $r$, which is the domain
  representative, is the element belonging to the $i$--$j$ interval
  for which $\Delta$ is minimum. Consistently with intuition, the
  dissimilarity score, $\Delta$, takes on small or large values if
  respectively many or few domain members are in contact with the
  representative. For a given number of domains, the optimal domain
  partitioning is the one that minimizes the sum of the $\Delta$
  scores for the domains.

  For a given number of domains, the consistency of the steered-MD and
  HiC-based subdivisions was measured by establishing a one-to-one
  correspondence of each domain in the two cases and next measuring
  the percentage of elements, $q$, having identical domain
  assignment. The one-to-one domain correspondence was identified by
  exploring the combinatorial space of correspondences and picking the
  one yielding the largest value of $q$.

% Do NOT remove this, even if you are not including acknowledgments
\section*{Acknowledgments}
We thank A. Pagnani and J. Nasica-Labouze for useful discussions and
G. Bussi for numerical advice. We acknowledge support from the Italian
Ministry of Education, grant PRIN 2010HXAW77.

%\section*{References}
% The bibtex filename
\bibliography{biblio}

\newpage

\section*{Figure Legends}

\setcounter{figure}{0}
\renewcommand{\thefigure}{S\arabic{figure}}

\begin{figure}[!ht]
\begin{center}
\includegraphics[width=4in]{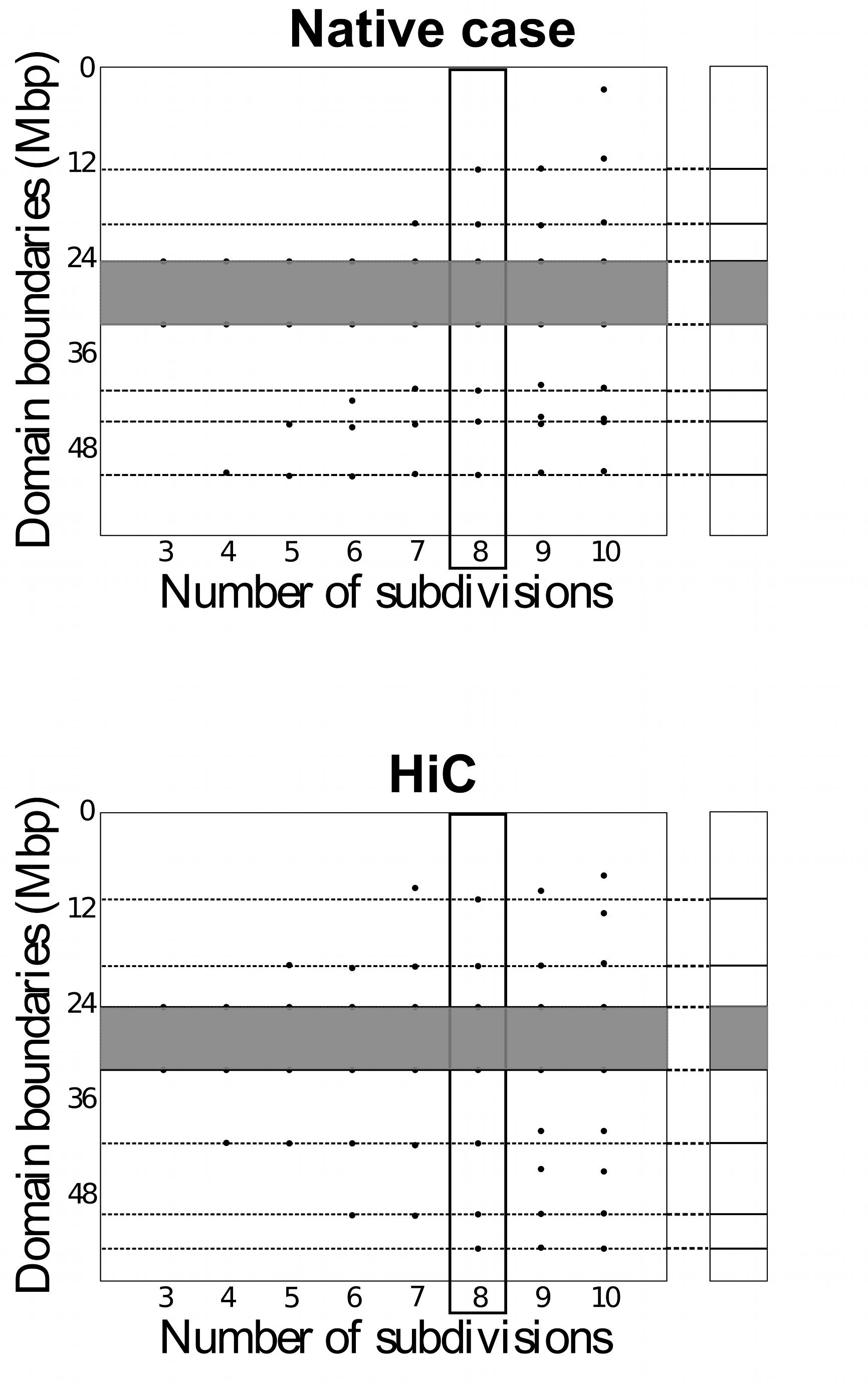}
\end{center}
\caption{ {\bf Chr19 spatial macrodomains.} The filled circles mark
  the boundaries of the Chr19 spatial macrodomains obtained from the
  clustering analysis of the steered-MD contact maps (top) and
  inferred from HiC data (bottom). The number of imposed macrodomains
  is shown on the $x$ axis. In all cases, one domain was fixed to
  correspond to the centromere (for which no HiC data are available)
  which is shown in grey. The dashed guidelines mark the subdivision
  into eight macrodomains which, by visual inspection provides robust,
  consensual boundaries in both cases. For claririty, the eight-domain
  subdivision is also reported on the chromosome sketch on the right.}
\label{fig:consensus_domains}
\end{figure}

\begin{figure}[!ht]
\begin{center}
\includegraphics[width=4in]{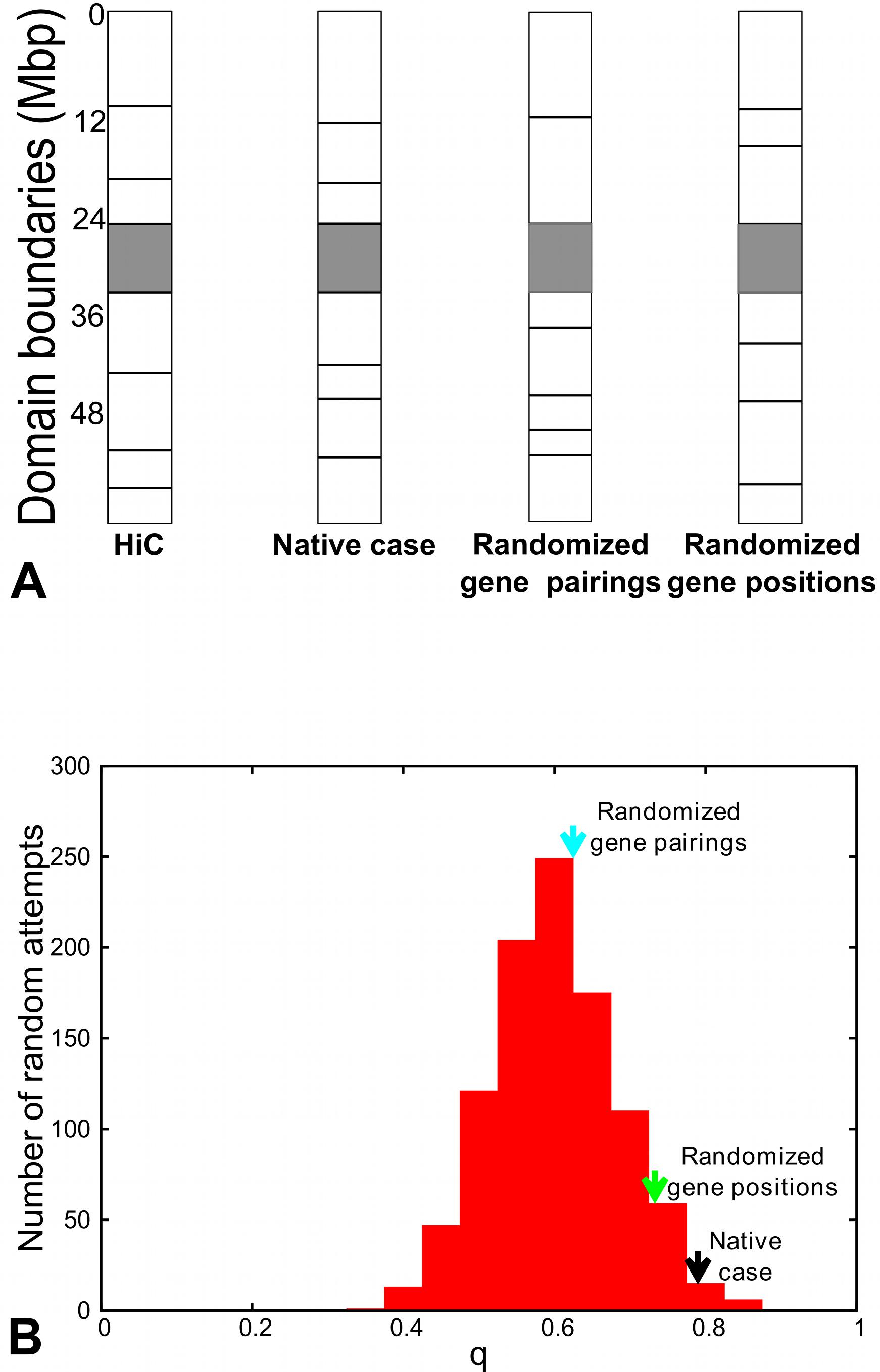}
\end{center}
\caption{ {\bf Comparison of macrodomain subdivisions.} (A). Schematic
  representation of the Chr19 partitioning in $8$ macrodomains (one
  being the centromere) based on the clustering analysis of contact
  maps inferred from HiC data and from steered-MD simulations on the
  native and randomized versions of the gene pairing network. In all
  cases, one domain was constrained to match the centromere (shown in
  grey). The overlap, $q$ and associated $p$-value of the steered-MD
  subdivisions against the reference HiC-data based one are as
  follows, (i) native case: $q=0.79$, $p$-value= 0.027; (ii)
  randomized gene positions: $q=0.73$, $p$-value=0.113; (iii)
  randomized gene pairings: $q=0.63$, $p$-value=0.49. The $p$-values
  were computed by comparing the observed overlap against a reference
  distribution of overlaps of $1000$ random chromosome partitions into
  $8$ domains (one always corresponding to the centromere). The
  reference distribution is shown in panel B. The arrows indicate the
  overlaps of the native and randomized cases.}
\label{fig:chrom_comparison}
\end{figure}

\begin{figure}[!ht]
\begin{center}
\includegraphics[width=4in]{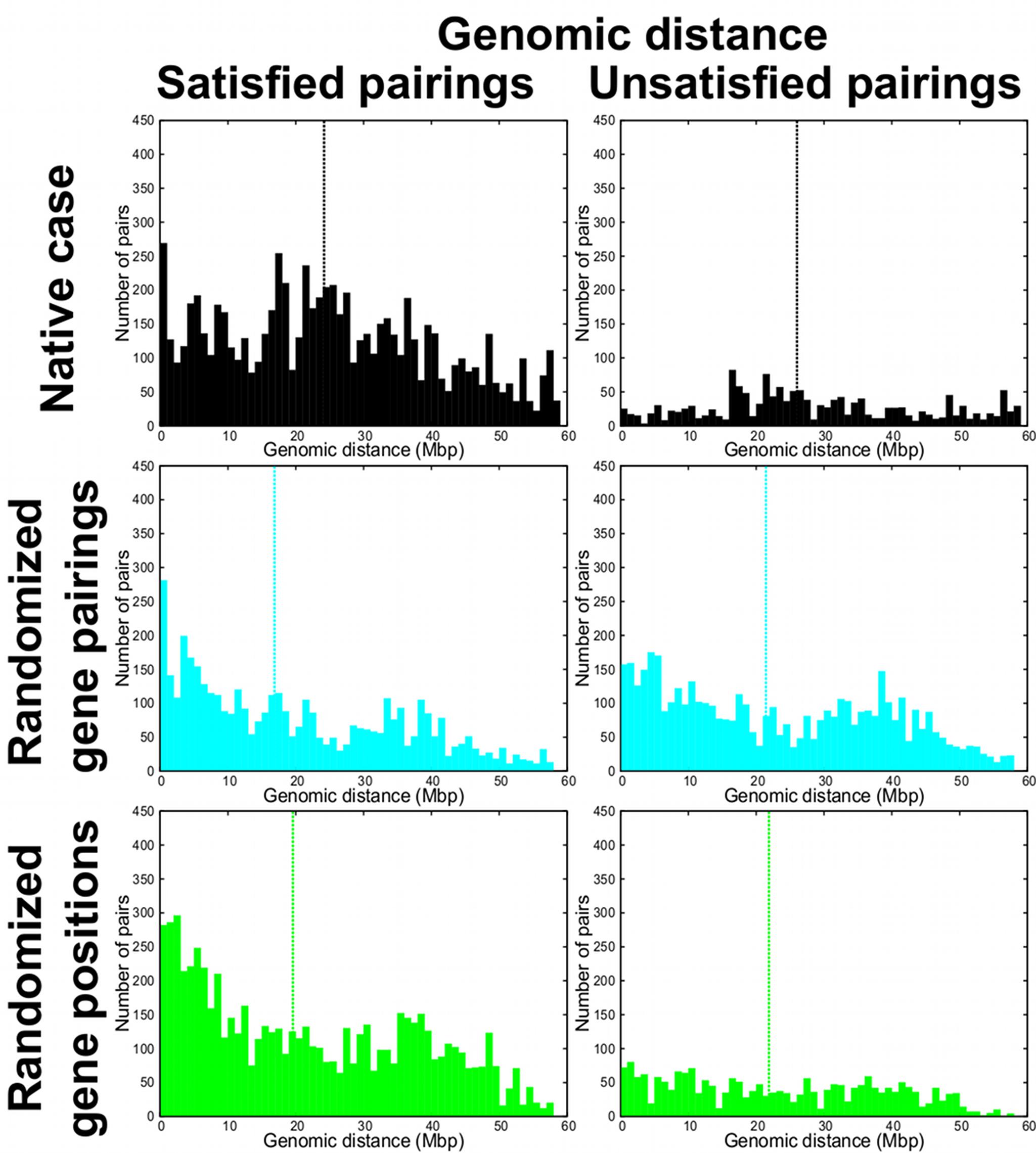}
\end{center}
\caption{ {\bf Genomic distance distribution for the target gene
    pairings established at the end of the steering protocol.} The
  plots on the left provide the genomic distance distributions of
  target gene pairings that are actually satisfied at the end of the
  steering protocols for the native and randomized cases. The
  analogous distribution for non-satisfied pairings is shown on the
  right. Dashed lines correspond to the median values. The results are
  cumulated over all $6$ chromosomes copies in the simulation box.}
\label{fig:unsatisfied_pairings}
\end{figure}

\begin{figure}[!ht]
\begin{center}
\includegraphics[width=4in]{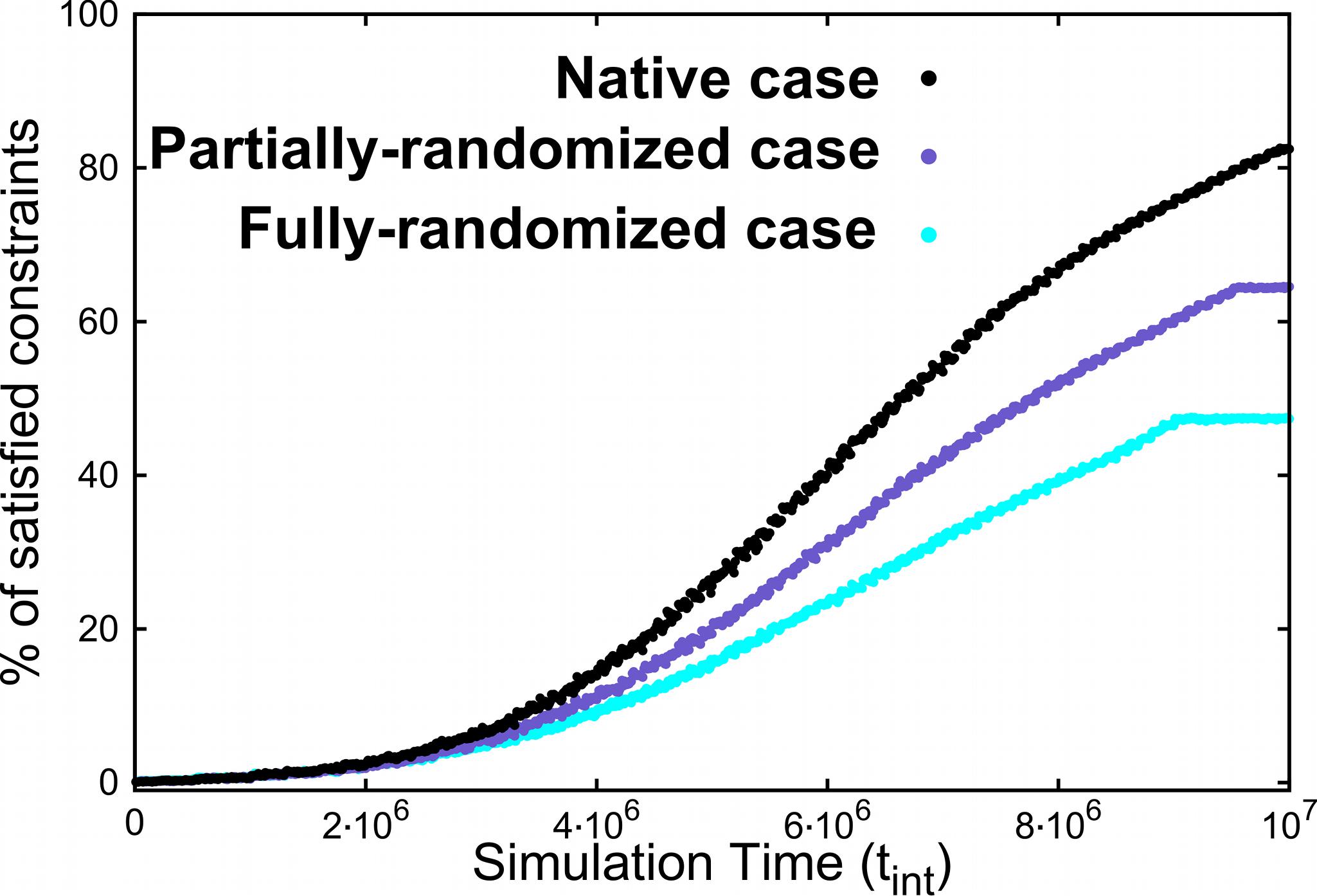}
\end{center}
\caption{ {\bf Gene colocalizability and gene network cliquishness.
    The time evolution of the fraction of satisfied gene pairings for
    three different steered-MD simulations.} The target gene pairing
  networks for the simulations are: the native network and two
  variants of it obtained by partial and full randomizations of gene
  pairings. The curves for the native and fully-randomized cases are
  the same as in Fig.~\ref{fig:IniVariantsResults}. The different
  cliquishness of the three target networks is captured by their
  clustering coefficient: $0.47$ for the native case, $0.30$ for the
  partially-randomized case and $0.12$ for the fully-randomized
  case. The fraction of established pairings shows a clear monotonic
  (increasing) dependence with the clustering coefficient.}
\label{fig:CC_overlap}
\end{figure}

%\section*{Tables}
%\begin{table}[!ht]
%\caption{
%\bf{Table title}}
%\begin{tabular}{|c|c|c|}
%table information
%\end{tabular}
%\begin{flushleft}Table caption
%\end{flushleft}
%\label{tab:label}
% \end{table}

\end{document}